\documentclass[prb,twocolumn,english,showpacs]{revtex4-1}
\usepackage{graphicx}
\usepackage{amssymb}
\usepackage{color}
\usepackage{times}

\begin{document}

\title{Attractive and repulsive Fermi polarons in two dimensions}

\author{Marco Koschorreck, Daniel Pertot, Enrico Vogt, Bernd Fr{\"o}hlich, Michael Feld, and Michael K{\"o}hl}
\affiliation{Cavendish\,Laboratory,\,University~of Cambridge, JJ Thomson Avenue, Cambridge CB30HE, United Kingdom}
\date{\today}

\maketitle

\textbf{The dynamics of a single impurity in an environment is a fundamental problem in many-body physics. In the solid state, a well-known case is an impurity coupled to a bosonic bath, for example lattice vibrations. Here the impurity together with its accompanying lattice distortion form a new entity, a polaron. This quasiparticle plays an important role in the spectral function of high-T$_c$ superconductors as well as in colossal-magnetoresistance in manganites\cite{Devreese2009}. For impurities in a fermionic bath, the attention so far has been mostly on heavy or immobile impurities which exhibit Anderson's orthogonality catastrophe\cite{Anderson1967} and the Kondo effect\cite{Kondo1964}. Only recently, mobile impurities have moved into the focus of research and they have been found to form new quasiparticles, so called Fermi polarons\cite{Svistunov2008,Schirotzek2009,Nascimbene2009,Kohstall2011}. The Fermi polaron problem constitutes the extreme, but conceptually simple, limit of two important quantum many-body problems: the BEC-BCS crossover with spin-imbalance\cite{Chevy2010} for attractive interactions and Stoner's itinerant ferromagnetism\cite{Duine2005} for repulsive interactions. It has been proposed that this and other yet elusive exotic quantum phases might become realizable in Fermi gases confined to two dimensions\cite{Chubukov1993,Hofstetter2002,Conduit2008}. Their stability and observability is intimately related to the theoretically debated\cite{Parish2011,Zollner2011,Klawunn2011,Schmidt2011,Ngampruetikorn2011} properties of the Fermi polaron in two dimensional Fermi gas. Here we create and investigate these Fermi polarons and measure their spectral function using momentum-resolved photoemission spectroscopy\cite{Dao2007,Stewart2008,Feld2011}. For attractive interactions we find evidence for the disputed pairing transition between polarons and tightly bound dimers, which provides insight into the elementary pairing mechanism of imbalanced, strongly-coupled two-dimensional Fermi gases. Additionally, for repulsive interactions we study novel quasiparticles, repulsive polarons, whose lifetime determine the possibility of stabilizing repulsively interacting Fermi systems.}

\begin{figure}
 \includegraphics[width=\columnwidth,clip=true]{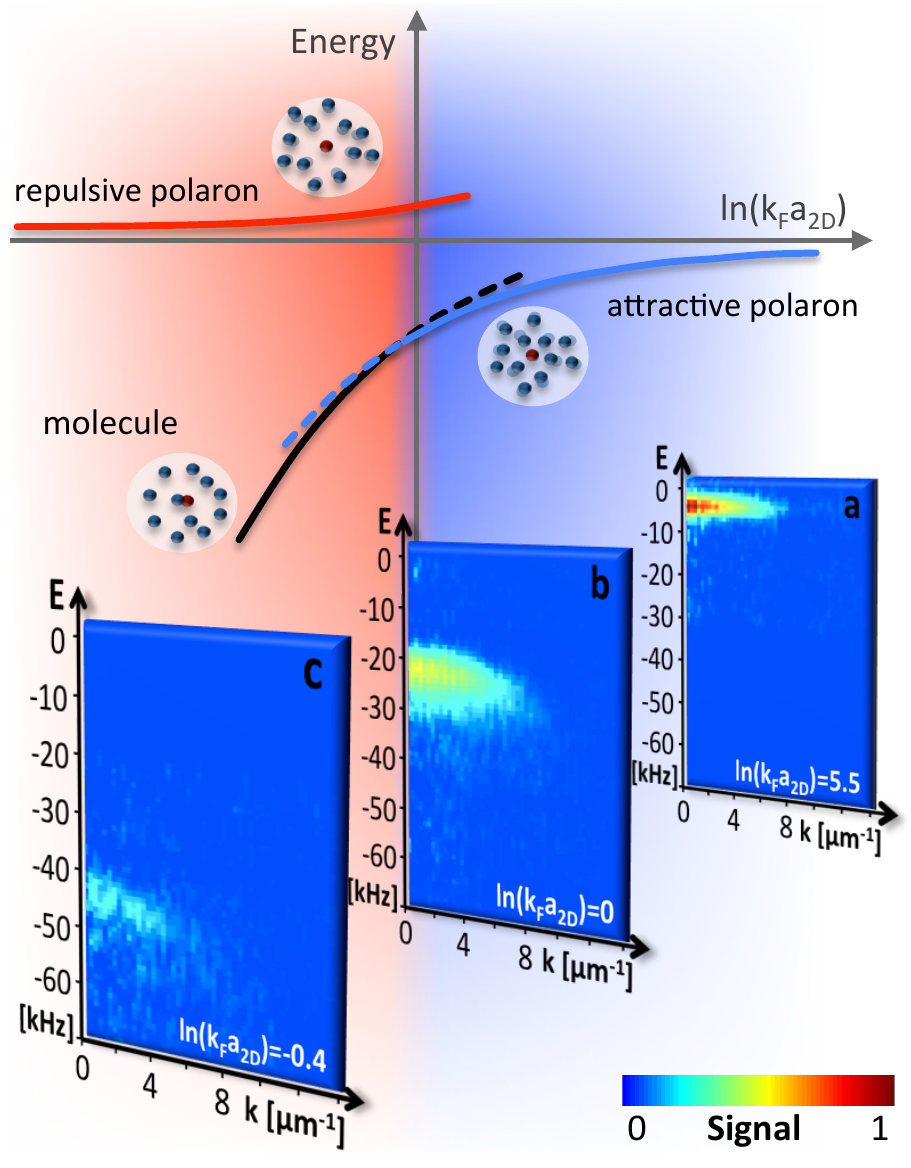}
 \caption{Polaron energy level diagram and measured single particle spectral function. For positive values of the interaction parameter $\ln(k_Fa_{2D})$ (see text) the attractive polaron is the quantum ground state. The single particle spectral function displays a sharp quasiparticle peak, \textbf{a}. For large negative values of $\ln(k_Fa_{2D})$ the ground state is molecular for which the single particle spectral function exhibits a weak incoherent feature, \textbf{c}. In between, \textbf{b}, there is a transition between the two limits. The repulsive polaron is a metastable state on the repulsive branch of the Feshbach resonance. In the spectral functions the free particle dispersion is implicitly subtracted.}
 \label{fig1}
\end{figure}

Spin-imbalanced Fermi gases\cite{Chevy2010} exhibit a wealth of complex pairing phenomena because the Fermi surfaces of the two spin components are mismatched. Famous examples are the predicted Fulde-Ferrell-Larkin-Ovchinikov (FFLO) pairing mechanism, in which Cooper pairs form at finite momentum, and the Chandrasekhar-Clogston limit of BCS superconductivity. In the extreme case of spin-imbalance, just one mobile spin-up particle is immersed in a large spin-down Fermi sea. New quasiparticles, so called Fermi polarons, are formed by the impurity being coherently dressed with particle-hole excitations of the majority component. For weak attractive interactions, the new many-body ground state is the attractive Fermi polaron (see Figure 1), which corresponds to the limit of a partially polarized Fermi gas. If the attraction between impurity and majority component is sufficiently strong, and depending on the dimensionality, the attractive polaron may undergo a transition to a locally paired dimer state\cite{Svistunov2008,Parish2011,Zollner2011}. This corresponds to the limiting case of phase separation between a fully polarized Fermi gas and a balanced superfluid state. The opposite situation of an impurity interacting repulsively with the fermionic bath is notably different. For short-range interactions, strong repulsion between particles can only be achieved if the underlying interaction potential is attractive which implies a two-particle bound state with binding energy $E_B$. Repulsive impurities are therefore metastable and eventually decay either into a bound state or into an attractive polaron under the simultaneous creation of particle and hole excitations. Despite its metastability, it has very recently been theoretically proposed\cite{Schmidt2011,Ngampruetikorn2011} that a repulsively interacting impurity still forms a well-defined quasiparticle, the repulsive polaron, which has not been observed in solid state systems so far. Its lifetime is of critical importance in pairing instabilities of repulsively interacting Fermi gases\cite{Pietila2011} and in the approach to observing the Stoner transition of itinerant ferromagnetism\cite{Duine2005}.

Dimensionality plays an important role in the physics of the Fermi polaron because quantum fluctuations, such as the creation of particle-hole pairs, are enhanced in low dimensions and affect the many-body ground state. In three dimensions, the properties of the Fermi polaron and the transition between polaron and molecule were predicted theoretically\cite{Lobo2006,Svistunov2008,Punk2009,Chevy2010,Pilati2010} and attractive\cite{Schirotzek2009,Nascimbene2009,Kohstall2011} as well as repulsive\cite{Kohstall2011} polarons have been observed experimentally. In one dimension, the exact solution of the spin-impurity Hamiltonian displays no transition for any interaction strength\cite{McGuire1966}. The most interesting, yet unresolved, case is in two dimensions: Different theoretical models\cite{Parish2011,Zollner2011,Klawunn2011,Schmidt2011,Ngampruetikorn2011} have predicted various scenarios for the quasiparticle properties and the existence or absence of a polaron-molecule transition on the attractive branch.

Here, we report on the creation of Fermi polarons by immersing a few spin-down impurity atoms in a two-dimensional Fermi sea of spin-up atoms. This strongly spin-imbalanced Fermi gas is formed by a 85/15 mixture of the two lowest Zeeman states $|-9/2\rangle$ and $|-7/2\rangle$ of the $F=9/2$ hyperfine manifold of $^{40}$K confined to two dimensions with an optical lattice\cite{Frohlich2011}. We adiabatically prepare the attractive ground state, and, independently, the repulsive branch, and measure their single particle spectral function $A(k,E)$ using momentum-resolved photoemission spectroscopy\cite{Dao2007,Stewart2008,Feld2011} (see Methods). Here, $k$ and $E$ are wave vector and energy of the single particle excitation, respectively. This methodology allows us to study the many-body state in equilibrium and is in contrast to ``inverse'' radio frequency (r.f.) spectroscopy\cite{Frohlich2011,Kohstall2011,Schmidt2011a} which probes the transient creation of excitations. Additionally, the single particle spectral function gives access to the full quasiparticle dispersion, which is averaged out in momentum-integrated r.f. spectroscopy\cite{Schirotzek2009,Sommer2012,Kohstall2011}. On the lower branch of the energy spectrum (see Figure 1 and Supplementary Material) we observe attractive Fermi polarons, measure their quasiparticle properties, and investigate the transition to the molecular state. We study the dependence of the quasiparticle properties on temperature, and, moreover, find evidence for quasiparticle interactions when we increase the impurity concentration. On the upper branch of the energy spectrum, we study repulsive Fermi polarons and, in particular, investigate their lifetime.

\begin{figure}
 \includegraphics[width=\columnwidth,clip=true]{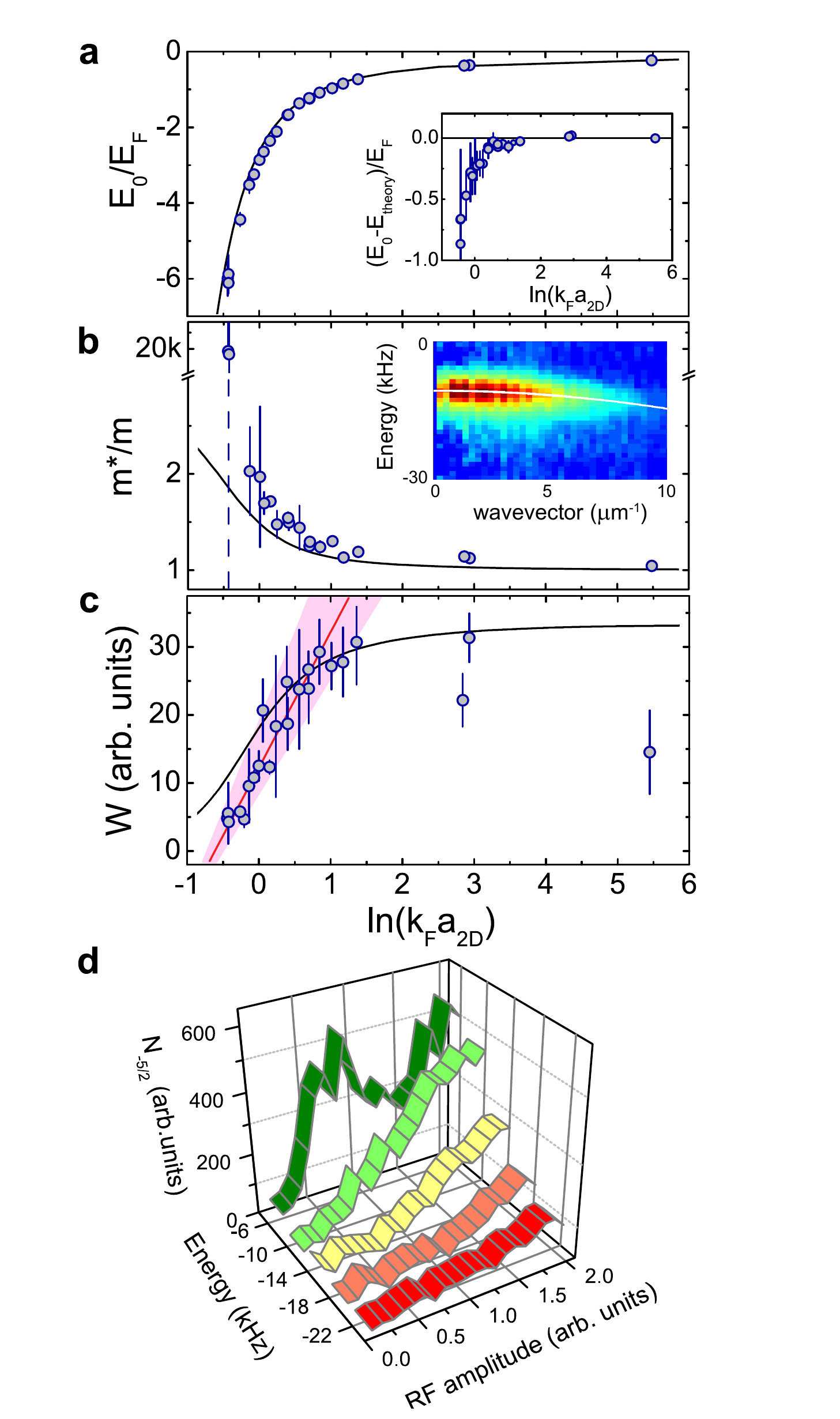}
 \caption{Attractive polaron. \textbf{a} Energy and \textbf{b} effective mass of the quasiparticle peak compared to the theoretical prediction\cite{Schmidt2011} (solid line). The dashed vertical line indicates the limit for reliably determining the effective quasiparticle mass. The inset in \textbf{a} shows the difference between experiment and theory. The inset in \textbf{b} shows an example of a fit used to determine the effective mass at $\ln(k_Fa_{2D})=0.7$. \textbf{c} Measured spectral weight and theoretical prediction for the quasiparticle weight\cite{Schmidt2011} (solid line). The red line signals the linear extrapolation and the shaded region the 1-$\sigma$ uncertainty. \textbf{d} Coherence of the quasiparticle at $\ln(k_Fa_{2D})=2.0(1)$. On the polaron resonance ($E/h=-6$\,kHz) we observe coherent oscillations whereas the incoherent background at higher frequencies ($E/h<- 6$\,kHz) increases linearly with the r.f. amplitude.}
 \label{fig2}
\end{figure}

In Figure 1a-c we display the measured single particle spectral function for different values of the interaction parameter $\ln(k_Fa_{2D})$. Here, $k_F$ denotes the Fermi wave vector of the majority component and $a_{2D}$ the two-dimensional scattering length (see Methods). For weak attractive interaction in the polaronic regime, we observe a symmetric line shape as a function of energy for every wave vector $k$, which signals a well defined quasiparticle, the attractive polaron (see Figure 1a). In the molecular regime we find a weak, asymmetric spectrum, which is skewed with a tail to lower energies (see Figure 1c). This spectrum is dominated by a broad incoherent background, which has contributions from incoherent particle-hole excitations and molecules.

In Figure 2 we display our measurements of the lower branch of the energy spectrum of the quasiparticle energy $E_0$ (a), effective mass $m^*$ (b), and spectral weight $W$ (c). We determine the energy $E_{peak}(k)$ of the maximum of the spectral function $A(k,E)$ at a wave vector $k$ and use the fit function $E_{peak}(k)=E_0+\hbar^2 k^2(1/m^*-1/m)/2$, with $m$ being the bare mass, in order to determine the quasiparticle energy $E_0$ and the effective mass $m^*$. The spectral weight $W$ is determined as the energy-integrated signal at $k=0$. We find general agreement with the theoretical prediction for the attractive polaron\cite{Schmidt2011} over a wide range of parameters. For data below $\ln(k_Fa_{2D})=-0.4$, we observe a divergence of the effective mass, which we interpret as the polaron-molecule transition. The polaron-molecule transition only has a very weak effect on the energy spectrum, as was already pointed out theoretically in three dimensions\cite{Svistunov2008}, since the polaron and molecule energy curves are crossing at a very shallow angle (see Figure 1). We observe a gradual deviation of the quasiparticle energy from theory (inset of Figure 2a) which, at least in part, could be attributed to finite range corrections to the interatomic potential. The spectral weight (Figure 2c) decreases continuously as $ln(k_Fa_{2D})$ is decreased [or $|E_0|$ is increased]. In order to extrapolate to the zero-crossing of the spectral weight, we employ a Monte-Carlo sampling of linear fits of all ranges of more than 3 data points and we select the subset of samples with a fit error below the median of the errors. The value for the zero-crossing obtained with this method is $ln(k_Fa_{2D})=-0.61(13)$ (see Figure 2c). In contrast, the theoretical quasiparticle weight (solid line, from ref.\cite{Schmidt2011}) evolves smoothly across the transition as it also accounts for metastable polaron quasiparticles in the regime where the molecule is the ground state. The polaron-molecule transition in two dimensions has been predicted\cite{Parish2011} at $\ln(k_Fa_{2D})=-0.8$, slightly lower than our observed value. Effects of the quasi-2D nature ($|E_0| \approx \hbar \omega$), averaging over the inhomogeneous density profile, final state interactions, finite temperature, and finite concentration of impurities may affect this value and shift the transition point to larger values of $\ln(k_Fa_{2D})$.

The attractive polaron is formed by a coherent dressing of the impurity with particle-hole excitations of the majority component. We investigate its coherence by driving Rabi oscillations between the polaron and the weakly interacting final state (Figure 2d). When the r.f. frequency is tuned to the polaron resonance in the spectrum, we observe coherent oscillations of the transferred number of atoms as we increase the amplitude of the r.f. coupling for a fixed duration. In contrast, when the r.f. frequency is tuned to the incoherent background we do not observe oscillations but only a linear increase in signal strength.


\begin{figure}
 \includegraphics[width=\columnwidth,clip=true]{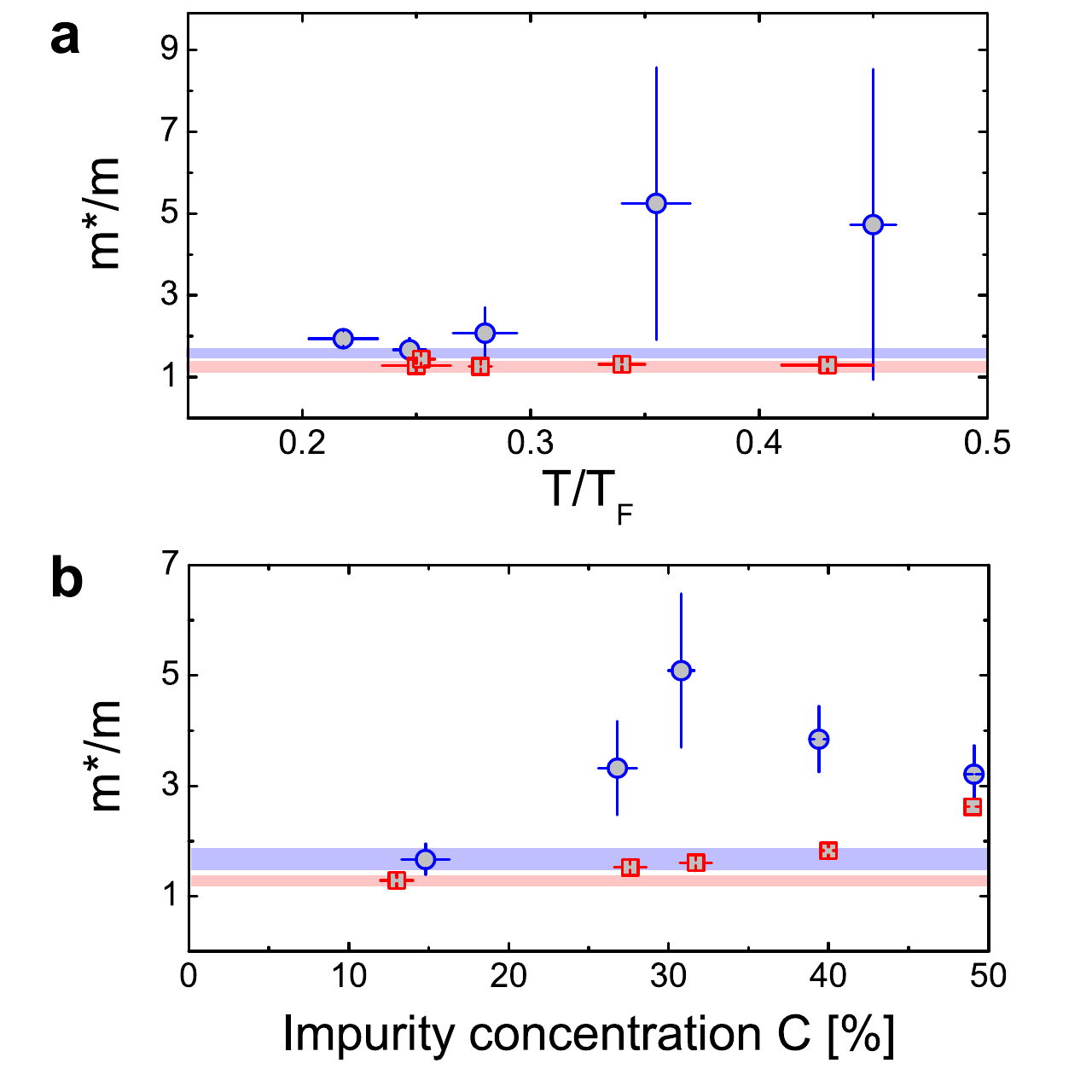}
 \caption{Finite temperature and impurity concentration. \textbf{a} Effective mass as a function of temperature for $\ln(k_Fa_{2D})= 0.45(5)$ (red squares) and $\ln(k_Fa_{2D})= -0.07(3)$ (blue circles). The solid horizontal lines show the theoretical prediction at zero temperature. All data are for a 87/13 mixture. \textbf{b} Effective mass as a function of impurity concentration. The horizontal lines show the theoretical prediction for the polaron and the shaded bars reflect the variation of the Fermi energy for the different measurements. Data are taken at $T/T_F=0.25(3)$.}
 \label{fig3}
\end{figure}

Only at low temperature and in the limit of a single impurity, polarons are well-defined quasiparticles. We investigate the deviations from this idealized picture (see Figure 3). Thermal excitations lead to decoherence of the dressing cloud and energetically higher lying states, such as the molecular branch (see Figure 1), may be occupied. In the weakly interacting limit, the molecular branch is approximately $E_F$ above the polaronic state, however, this energy difference closes down as one approaches the polaron-molecule transition. Hence one expects the effects of temperature to be most pronounced near the polaron-molecule transition and we focus our investigations on this regime. Indeed, we find that at $\ln(k_Fa_{2D})=0.45(5)$ (red squares) the effective mass remains unaffected by temperature up to $T/T_F=0.5$  (see Figure 3a) whereas for $\ln(k_Fa_{2D})=-0.07(3)$ (blue circles) we observe a significant change of the single particle spectrum. At low temperatures we find good agreement with the zero-temperature prediction but at higher temperatures the effective mass increases. This could signal a transition from a polaron at low temperature into the incoherent molecule-hole continuum at high temperature. Another deviation from the idealized polaron picture results from a finite concentration of impurities. This may cause interactions between quasiparticles mediated by the majority species. In Figure 3b we show the dependence of the quasiparticle properties on the impurity concentration $C=N_\uparrow/(N_\uparrow+N_\downarrow)$ of the Fermi gas. In the limit $C\ll1$, the observed effective mass matches well with the theoretical prediction for the polaron. Upon increasing the impurity concentration, we find that the effective mass of the polarons increases, which we attribute to polaron-polaron interaction. Again, we find that this effect is more pronounced near the polaron-molecule transition. At equal population we observe approximately the same effective mass for both interaction strengths.


Finally, we turn our attention to the repulsive polaron. This quasiparticle plays a decisive role in the question, whether itinerant ferromagnetism can be achieved with repulsive short-range interactions between fermions, which has been argued to be impossible in a three-dimensional spin-1/2 mixture with contact interactions\cite{Sanner2011}. The recent observation of a repulsive polaron in a three-dimensional heteronuclear Fermi-Fermi mixture\cite{Kohstall2011} has renewed the interest in whether mass-imbalance and/or reduced dimensionality could stabilize the formation of polarized domains. We access the repulsive branch of the Feshbach resonance between the $|-9/2\rangle$ and $|-5/2\rangle$ state near 224\,G (see Methods). Lifetime, energy, and effective mass are extracted from the spectral function $A(k,E)$, which we measure after different hold times. For the lifetime measurement (Figure 4a), we determine the atom number loss, integrating from -10\,kHz to +10\,kHz. We observe a fast decay (inset of Figure 4a) to a plateau which decays much more slowly, similar to the balanced case in three dimensions\cite{Sanner2011}. In Figure 4a we display the measured lifetime of the repulsive branch as a function of the interaction strength, which agrees with the theoretical prediction for the lifetime of the repulsive polaron\cite{Ngampruetikorn2011}, adapted for the quasi-2D binding energy $E_B$ and quasi-2D scattering length $a_{2D}$ (see Methods). In the regime of strong interactions (small negative $\ln(k_Fa_{2D})$) the measured lifetime becomes a sizeable fraction of the inverse Fermi energy $h/E_F=0.1\,$ms, qualitatively similar to repulsive Fermi gases in three dimensions\cite{Sanner2011}. The energy and the effective mass are extracted from the spectral function as above and they are plotted in Figures 4b and 4c, respectively. Both mass and energy deviate from the prediction for the homogeneous gas under exact two-dimensional confinement\cite{Schmidt2011,Ngampruetikorn2011}, for which we replace the exact-2D binding energy with the quasi-2D binding energy. The repulsive polaron energy is in qualitative agreement with the trap-averaged spectra of ref.\cite{Schmidt2011}, which have revealed a relatively small dependence of the polaron energy on the interaction parameter. The inclusion of final-state interaction effects and a quasi-2D theory including harmonic confinement as well as possible effects of non-adiabaticity and the formation time of the polaron during the relatively fast magnetic field ramp could be required to quantitatively model our data.

\begin{figure}
 \includegraphics[width=\columnwidth,clip=true]{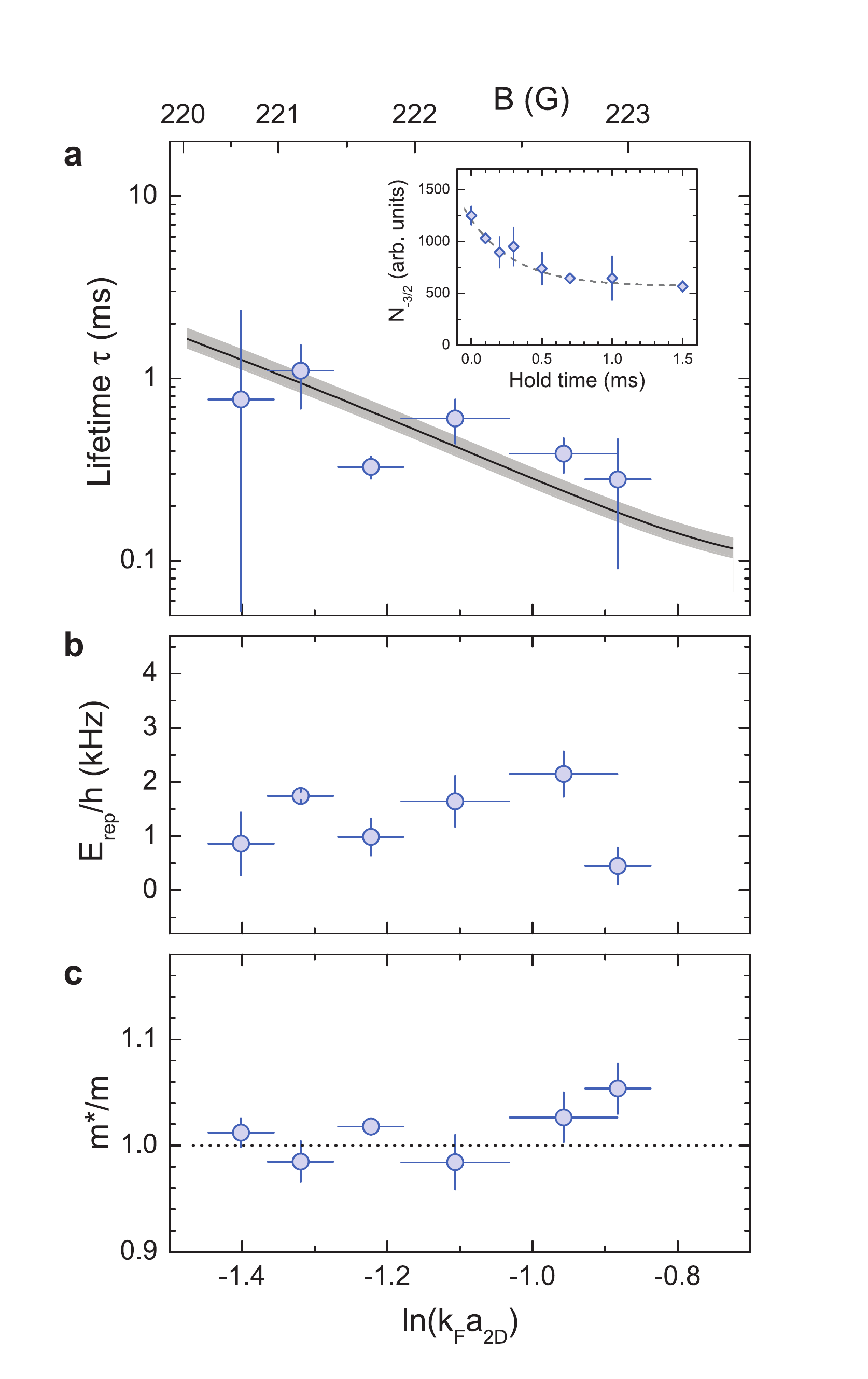}
 \caption{Repulsive polaron. \textbf{a} Lifetime of the repulsive branch. The inset shows the example of a time-resolved measurement at $\ln(k_Fa_{2D})=-1.2$ from which we extract the lifetime by a fit (dashed line). The solid line is the theoretical prediction from ref.\cite{Ngampruetikorn2011}. The grey shaded area reflects our range of Fermi energies of $(11\pm1)$\,kHz. Measured energy (\textbf{b}) and effective mass (\textbf{c}).}
 \label{fig4}
\end{figure}

In conclusion, we have observed features of attractive and repulsive Fermi polarons in two dimensions and find evidence for the polaron-molecule transition in a two-dimensional Fermi gas, yet at a slightly different interaction parameter than theoretically predicted. The future inclusion of a localized, potentially spin-dependent, potential will allow to increase the effective mass of the impurity and to protrude into the regime of x-ray edge phenomena and the Kondo effect, in which the internal spin degree of freedom of an immobile impurity becomes entangled with a Fermi sea.

\section*{Methods summary}

The starting point of this work is a quantum degenerate Fermi gas of $^{40}$K atoms in a strongly imbalanced 85/15 mixture of the two lowest Zeeman states $|-9/2\rangle$ and $|-7/2\rangle$  ($|-5/2\rangle$ for the repulsive polaron) of the $F=9/2$ hyperfine manifold confined to an optical lattice\cite{Frohlich2011}. The average Fermi energy of the majority component is $E_F/h= 10$\,kHz  and the temperature is typically $T/T_F=0.25$. We record momentum-resolved photoemission spectra near the Feshbach resonance of 202.1\,G (224.2\,G) by coupling the $|-7/2\rangle$ ($|-5/2\rangle$) state to the weakly interacting state $|-5/2\rangle$ ($|-3/2\rangle$) using an r.f. photon of frequency $\Omega_{rf}$ with negligible momentum transfer. The momentum distribution of the transferred atoms is detected in a time-of-flight experiment and we average the absorption signal azimuthally to obtain the single particle spectral function $A(k,E)$, with the wave vector $k=\sqrt{k_x^2+k_y^2}$ and energy $E=E_Z-\hbar \Omega_{rf}$ using the atomic Zeeman energy $E_Z$.

\section{Methods}

\subsection{Preparation of imbalanced 2D gases and momentum-resolved photoemission spectroscopy}
We prepare degenerate Fermi gases following the general procedure of ref.\cite{Frohlich2011}. Additionally, during the evaporative cooling sequence we apply an r.f. pulse in order to prepare the desired spin mixture by selectively removing atoms from the $|-7/2\rangle$ state. Subsequently, we load the imbalanced mixture into an optical lattice formed by a horizontally propagating, retro-reflected laser beam of wavelength $\lambda=1064$\,nm, focussed to a waist (1/e$^2$ radius) of 140\,$\mu$m. We increase the laser power over a time of 200\,ms to reach the final potential depth with a trapping frequency along the strongly confined direction of $\omega=2 \pi \times 78.5$\,kHz. After loading the optical lattice, we adiabatically reduce the power of the optical dipole trap such that the atoms are confined only by the Gaussian intensity envelope of the lattice laser beams. The resulting radial trapping frequency of the two-dimensional gases is $\omega_\perp=2\pi\times 127$\,Hz. Along the axial direction we populate approximately 30 layers of the optical lattice potential with an inhomogeneous peak density distribution. The Fermi energy (of the majority component) is the measured, density-weighted average across the different layers. We measure that majority and minority component have the same absolute temperature. As a result, the minority atoms occupy fewer layers and experience a more uniform peak density of majority atoms.

In order to access the attractive polaron branch, we adiabatically increase the interaction strength by lowering the magnetic field in 30\,ms from 209\,G to a value near the Feshbach resonance at 202.1\,G. Momentum resolved photoemission spectroscopy is performed by applying a radio-frequency pulse near 50\,MHz with a Gaussian amplitude envelope with a full width at half maximum of 280\,$\mu$s to transfer atoms from the $|-7/2\rangle$ state to the $|-5/2\rangle$ state. Atoms in the $|-5/2\rangle$ state have a two-body s-wave scattering length of 250 Bohr radii with the $|-9/2\rangle$ state\cite{Stewart2008}. We turn off the optical lattice 100\,$\mu$s after the radiofrequency pulse, switch off the magnetic field, and apply a magnetic field gradient to achieve spatial splitting of the three spin components in a Stern-Gerlach experiment.

The repulsive polaron branch is accessed by preparing the imbalanced $|-9/2\rangle/|-7/2\rangle$ mixture at a magnetic field of 220\,G and spin-flip the atoms from $|-7/2\rangle$ to $|-5/2\rangle$ using a 30\,$\mu$s long rf pulse with $95\%$ transfer efficiency. The pulse is tuned to the bare Zeeman transition energy but is broadband to include also the repulsive polaron energy shift. Then, we ramp the magnetic field to the desired value near the $|-9/2\rangle/|-5/2\rangle$ Feshbach resonance in 5\,ms. We apply a radio-frequency pulse near 53\,MHz with a Gaussian amplitude envelope with a full width at half maximum of 270\,$\mu$s to transfer atoms from the $|-5/2\rangle$ state to the $|-3/2\rangle$ state. The scattering length between the $|-9/2\rangle$ and $|-3/2\rangle$ states at this magnetic field is approximately 190\,$a_B$. The sequence is completed using absorption imaging of the expanding clouds after Stern-Gerlach separation in an inhomogeneous field.

For each experimental run, the magnetic field is calibrated using spin-rotation with an rf pulse of an imbalanced mixture on the $|-9/2\rangle$/$|-7/2\rangle$ transition. The magnetic field accuracy deduced from these measurements is $<3$\,mG. We measure the temperature by ballistic expansion of a weakly interacting gas, and the quoted numbers refer to the average of $T/T_F$ across the whole sample.

\subsection{Interaction parameter in two dimensions}
Two particles in a two-dimensional system created by tight harmonic confinement exhibit a confinement-induced bound state of binding energy $E_B$. The binding energy is linked to the three-dimensional scattering length $a_s$ and is obtained from the transcendental equation\cite{Petrov2001,Bloch2008}
\begin{equation}
l_z/a_s=\int_0^\infty \frac{du}{\sqrt{4 \pi u^3}} \left(1- \frac{\exp(-E_B u/(\hbar \omega))}{\sqrt{(1-\exp(-2 u))/(2 u)}}\right).
\end{equation}
Here, $l_z=\sqrt{\hbar/m\omega}$ and $a_s$ is determined using the following parameters of the Feshbach resonances: Feshbach resonance between $|-9/2\rangle$ and $|-7/2\rangle$: $B_0=202.1$\,G, $\Delta B=7$\,G and $a_{bg}=174\,a_B$ where $a_B$ is the Bohr radius, and Feshbach resonance between $|-9/2\rangle$ and $|-5/2\rangle$: $B_0=224.2$\,G, $\Delta B=7.5$\,G and $a_{bg}=174\,a_B$. Using $E_B$ we define the two-dimensional scattering length $a_{2D}$ by $E_B=\hbar^2/ma_{2D}^2$.

We thank S. Baur, N. Cooper, E. Demler, T. Enss, J. Levinsen, C. Kollath, M. Parish, R. Schmidt, and W. Zwerger for discussion and J. Bohn for communicating the unpublished details of the $|-5/2\rangle/|-3/2\rangle$-Feshbach resonance in $^{40}$K. The work has been supported by {EPSRC} (EP/G029547/1), Daimler-Benz Foundation (B.F.), Studienstiftung, and DAAD (M.F.).

The authors declare that they have no competing financial interests.

The experimental setup was devised and constructed by M.F., B.F., E.V., and M. K{\"o}hl, data taking was performed by M.K., E.V., and D.P., data analysis was performed by M.K. and D.P., and the manuscript was written by M. K{\"o}hl with contributions from all coauthors.

Correspondence and requests for materials should be addressed to M. K{\"o}hl~(email: mk540@cam.ac.uk).


\end{document}